\documentclass[RNAAS]{aastex62}

\begin{document}

\title{Unidentified FRBs in archival data}

\correspondingauthor{E. F. Keane}

\author[0000-0002-4553-655X]{E. F. Keane}
\affiliation{SKA Organization, Jodrell Bank Observatory, Macclesfield, Cheshire SK11 9DL, UK}
\email{ekeane@skatelescope.org}

\author[0000-0003-1301-966X]{D. R. Lorimer}
\affiliation{West Virginia University, Department of Physics and Astronomy, P. O. Box 6315, Morgantown, WV, USA}
\affiliation{Center for Gravitational Waves and Cosmology, West Virginia University, Chestnut Ridge Research Building, Morgantown, WV, USA}

\author[0000-0002-2578-0360]{F. Crawford}
\affiliation{Department of Physics and Astronomy, Franklin and Marshall College, PO Box 3003, Lancaster, PA 17604, USA}


\section{FRB 010312}

Recently \citet{Zhang2019} reported the discovery of FRB~010312. This event occurred during a survey of the Magellanic Clouds and surrounding regions wherein the first fast radio burst, FRB~010724, was discovered~\citep{Lorimer2007}. The reported signal-to-noise ratio (S/N) of FRB~010312 is $11$, for a pulse width of $24(1)$~ms. Examining the data\footnote{The relevant survey pointing is SMC018\_034.} with \textsc{destroy}\footnote{\texttt{https://github.com/evanocathain/destroy\_gutted}}, and performing no RFI mitigation, we confirm the burst detection and obtain a S/N$=12$ with a pulse width of $37^{+17}_{-8}$~ms, at the reported dispersion measure. This implies a peak flux density of $\sim 150$~mJy and specific fluence of $\sim 5.6$~Jy\,ms if the signal occurred at boresight. As Zhang et al. point out, their detection brings this survey more in-line with other expectations for the FRB rate~\citep{Champion2016,SUPERB2}.

\section{Why Missed?}

Much more interesting than the above minutiae is why and how this event was missed in the previous analyses of these data. The survey was originally searched~\citep{Crawford2016} using \textsc{seek} from the \textsc{sigproc}\footnote{\texttt{http://sigproc.sourceforge.net/}} software suite. This algorithm uses down-sampling and smoothing processes in searching a time series, that can result in under-estimated S/N values; these steps were initially motivated as a time saving measure to overcome computational constraints that now no longer exist. The upshot is that the S/N values reported by \textsc{seek} can be degraded by as much as a factor of $\sqrt{2}$~\citep{kp15}. For FRB~010312, this exact scenario occurred and the detection S/N determined was $8$ and based on this the candidate was dismissed as sub-threshold.

\section{Re-processing Archival Datasets}

Several archival datasets have only been searched using this and other pipelines that perform similarly or worse. Without re-processing these data with something optimised\footnote{For example \textsc{heimdall}, a GPU-based single pulse search code --- \texttt{https://sourceforge.net/projects/heimdall-astro/}}, in the worst case the true detectable population of FRBs in a dataset could, in a Euclidean Universe, be $2^{3/4}\approx 1.65$ times larger than one might initially estimate, i.e. a $10\sigma$ threshold would in fact be a $14\sigma$ threshold. Reprocessing these same data sets in search of radio pulsars as ever more sophisticated search techniques are developed, as RFI mitigation procedures improve and as computational constraints disappear has been successful over and over again (see \citealt{einstein} and references therein). This was, of course, true for the original FRB discovery and the same is likely true for FRB searches generally. As well illustrated by FRB~010312, there is potentially a large number of missed FRBs in datasets that have already been searched. The situation may even be more stark when we consider how difficult it is to mitigate radio frequency interference and that all FRB searches ever performed have looked for temporally symmetric broadband flat-spectrum pulses, and most FRBs do not look like that. We encourage further work using multiple search codes, where possible, and also searching a broad DM and pulse width range.

\acknowledgements

EFK would like to thank everybody who took part in FRB2019 in Amsterdam for a highly informative and enjoyable meeting. Endless reprocessing of FRB datasets is encouraged!

\begin{samepage}
\bibliographystyle{aasjournal}
\bibliography{rnaas_frb010312}
\end{samepage}


\end{document}